\documentclass[aps,prl,twocolumn,groupedaddress,showpacs]{revtex4}
	
\usepackage{graphicx,amsmath}
\usepackage{feynmp}
\usepackage{mathrsfs}
\usepackage{gensymb}
\usepackage{subfigure}
\usepackage{float}
\usepackage{setspace}
\usepackage{amssymb}
\usepackage{bm}
\usepackage{soul}

\usepackage{color}

\begin{document}

\title{Signatures of magnetic-field-driven quantum phase transitions in the entanglement entropy and spin dynamics of the Kitaev honeycomb model}

\author{David C. Ronquillo, Adu Vengal, and Nandini Trivedi}
\affiliation{Department of Physics, The Ohio State University, Columbus, Ohio 43210, USA}

\begin{abstract}
The main question we address is how to probe the fractionalized excitations of a quantum spin liquid (QSL), for example, in the Kitaev honeycomb model. By analyzing the energy spectrum and entanglement entropy, for antiferromagnetic couplings and a field along either [111] or $[001]$, we find a gapless QSL phase sandwiched between the non-Abelian Kitaev QSL and polarized phases. Increasing the field strength towards the polarized limit destroys this intermediate QSL phase, resulting in a considerable reduction in the number of frequency modes and the emergence of a beating pattern in the local dynamical correlations, possibly observable in pump-probe experiments. 
\end{abstract}

\date{\today}


\maketitle
\setstretch{1}

{\it Introduction.}
Signatures of the exotic fractionalized excitations of the two-dimensional Kitaev model on a honeycomb lattice \cite{Kitaev2006} have recently been of interest given their experimental accessibility within candidate Kitaev-like materials \cite{Singh2010, Singh2012, Plumb2014}. The exactly solvable Kitaev model consists of $S=1/2$ degrees of freedom that are frustrated by anisotropic bond-dependent, nearest-neighbor interactions
\begin{equation}\label{kit}
H_{K}=\sum_{\alpha}\left(J_{\alpha}\sum_{\langle j,k\rangle_{\alpha}}\sigma_{j}^{\alpha}\sigma_{k}^{\alpha}\right),
\end{equation}
where $J_{\alpha}$ is the Kitaev exchange constant, $\alpha\in\{x,y,z\}$, $j$ and $k$ are nearest-neighbor sites lying along the bond $\alpha$, and $\sigma_{j}^{\alpha}$, $\sigma_{k}^{\alpha}$ are the corresponding Pauli matrices. Recent theoretical advances reveal the identifying characteristics of fractionalized excitations in the dynamical structure factor \cite{Knolle2014, Knolle2015}, in a two-peak structure of the temperature-dependent entropy \cite{Nasu2015}, and in the longitudinal and transverse thermal conductivities \cite{Nasu2017}. These advances shed light on the nature of the Kitaev quantum spin liquid (QSL), whose fractionalized excitations within the gapped non-Abelian phase may find applications in possible quantum computing devices \cite{Nayak2008}.

Originally conceived as a toy model, researchers have gone on to consider the microscopic mechanisms necessary for realizing Kitaev physics in real materials. This has led to proposals of Hamiltonians with extended Kitaev-Heisenberg interactions \cite{Jackeli2009, Chaloupka2010, Chaloupka2013}, as well as additional symmetric off-diagonal interactions \cite{Rau2014, Winter2016}. Candidate compounds $\alpha$-RuCl$_{3}$ \cite{Banerjee2016, Banerjee2017}, and A$_{2}$IrO$_{3}$ (A = Na, Li) \cite{Mehlawat2017} show salient features in experiments that can be attributed to residual fractional excitations of the pure Kitaev phase proximate to these materials' zig-zag ordered ground state. Recently, experimental measurements of the thermal Hall conductivity $\kappa_{xy}$ of $\alpha-$RuCl$_{3}$ have revealed signatures of itinerant Majorana excitations in the sign, magnitude, and temperature ($T$) dependence of $\kappa_{xy}/T$ within $T_{N}=7$ K$<T<80$ K$\sim J_{\alpha}/k_{b}$, where $T_{N}$ is the temperature at which the zigzag order begins \cite{Kasahara2018}. Nuclear magnetic resonance (NMR) studies have revealed a cubic magnetic field dependence of the spin-excitation gap, consistent with Kitaev's prediction \cite{Kitaev2006} for Majorana fermions \cite{Jansa2018}.

\begin{figure*}[htb!]
\begin{center}
\includegraphics[scale = .135]{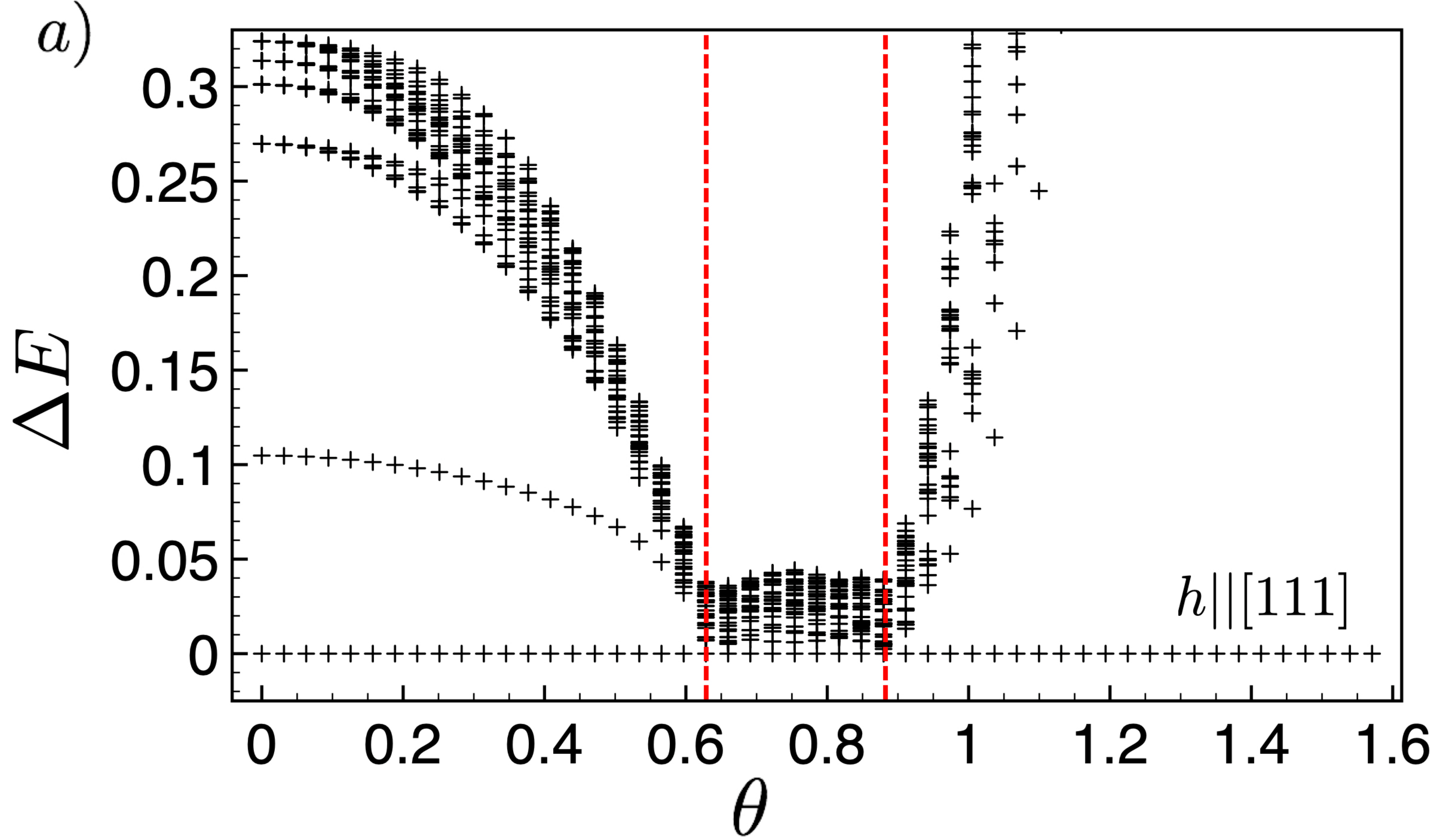}
\includegraphics[scale = .135]{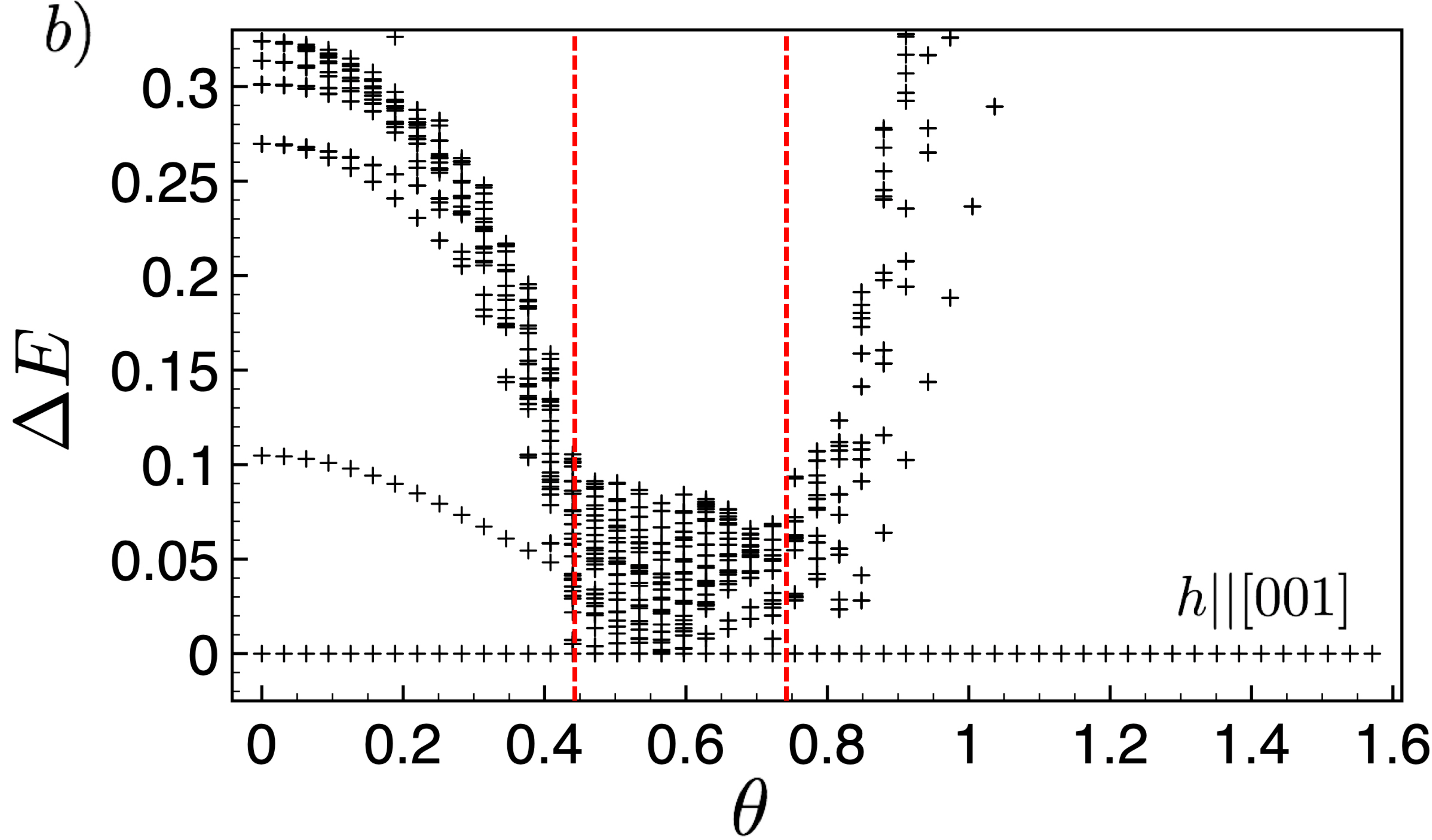}
\includegraphics[scale = .135]{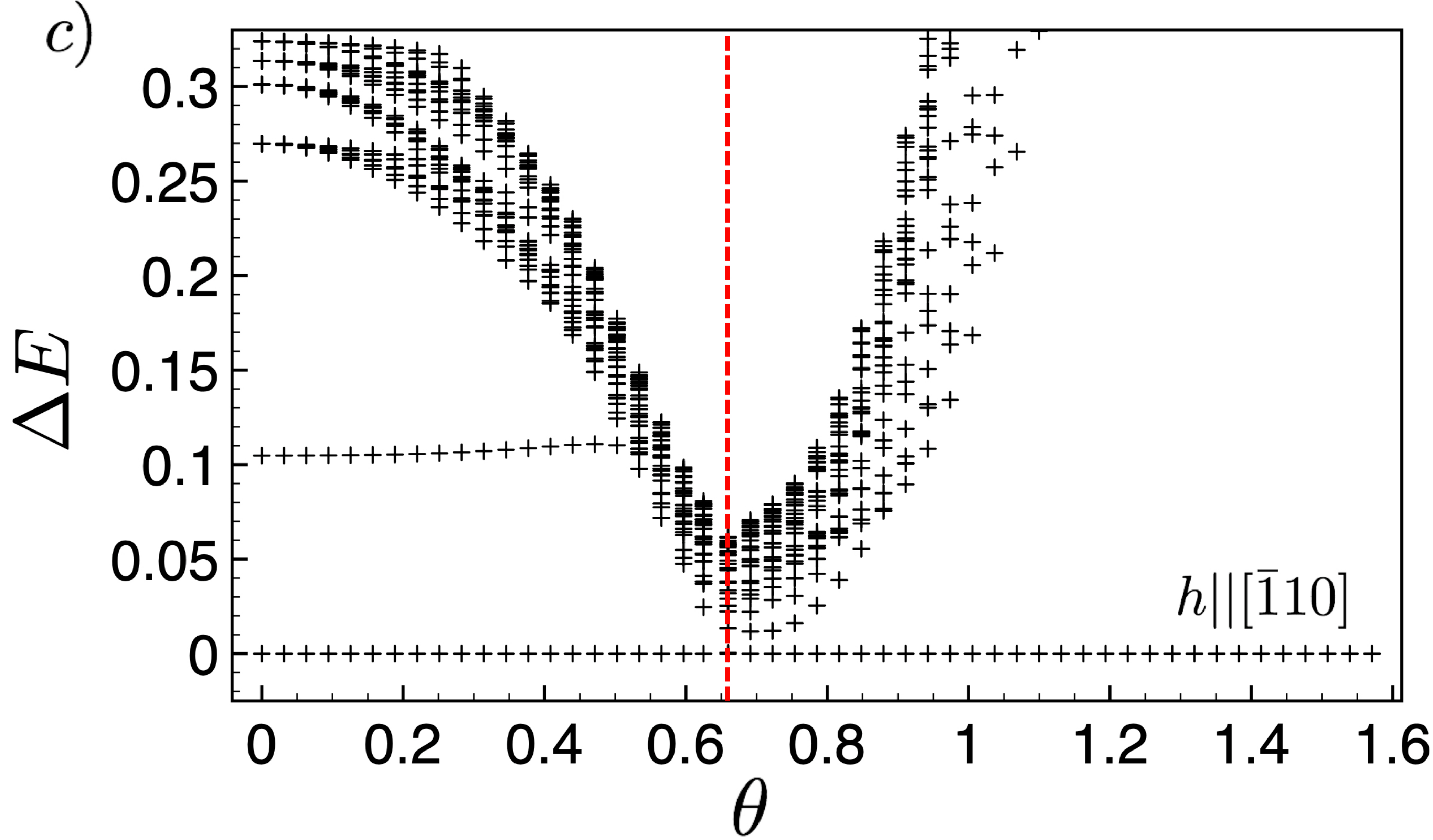}
\includegraphics[scale = .135]{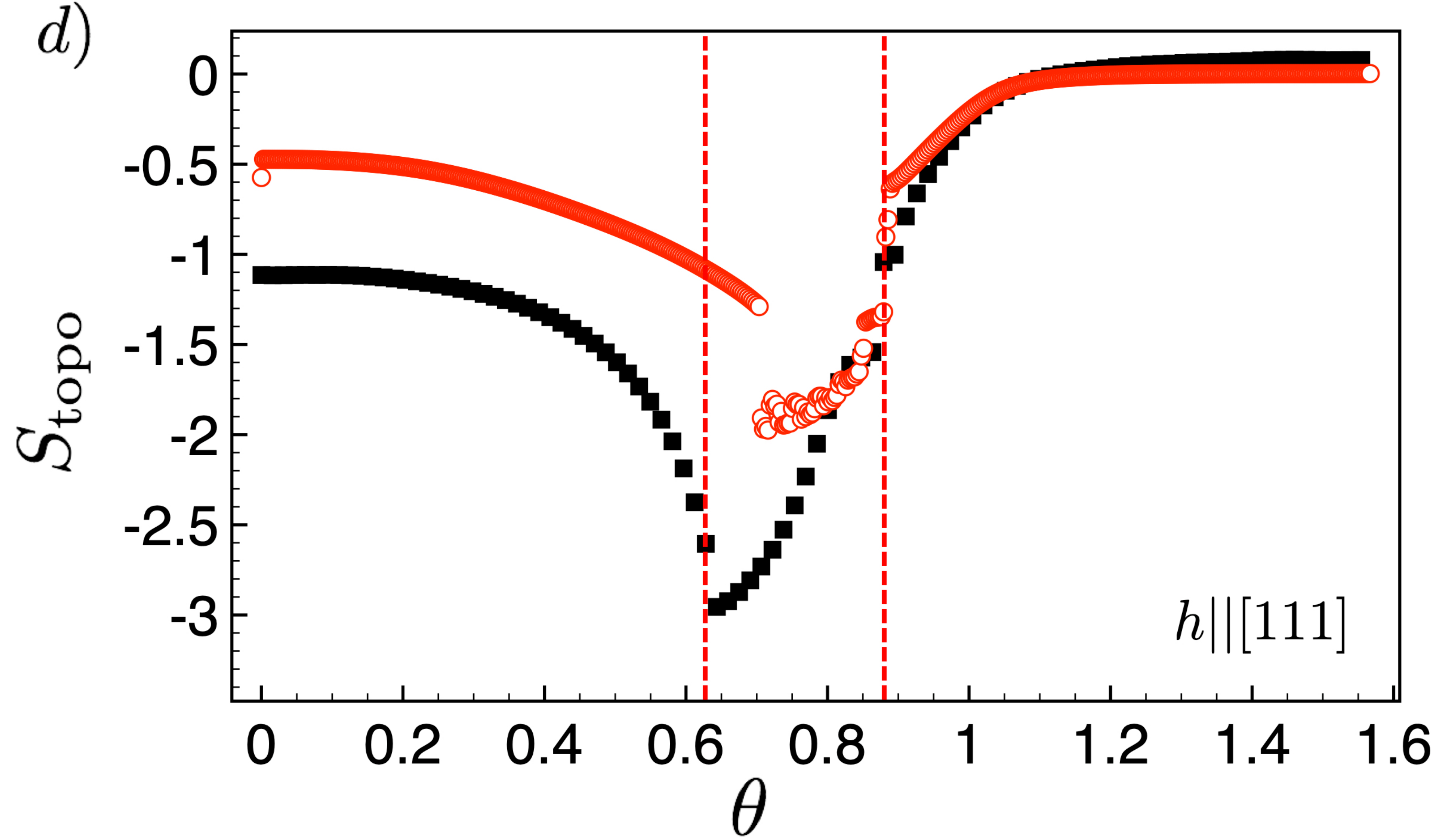}
\includegraphics[scale = .135]{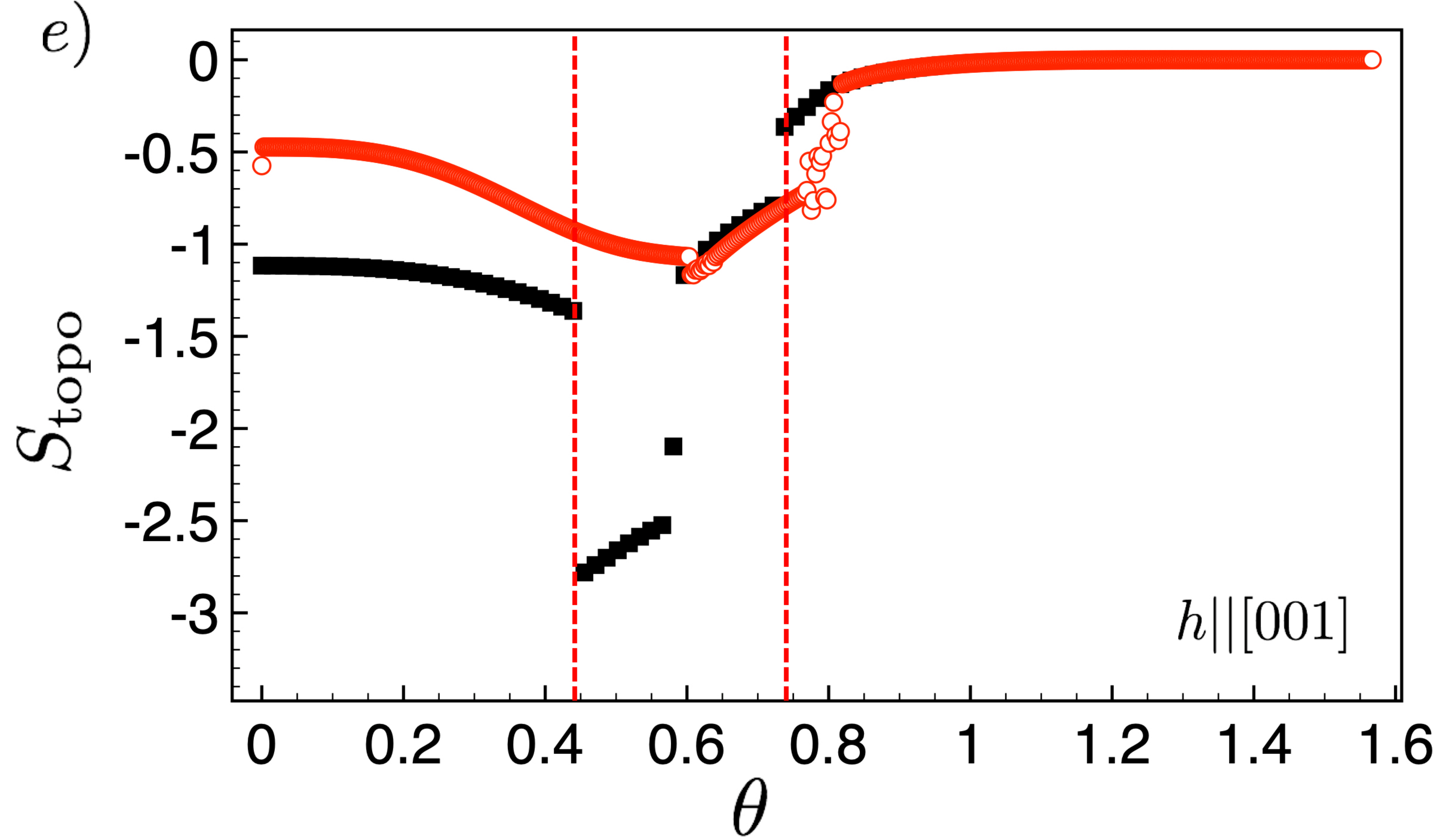}
\includegraphics[scale = .135]{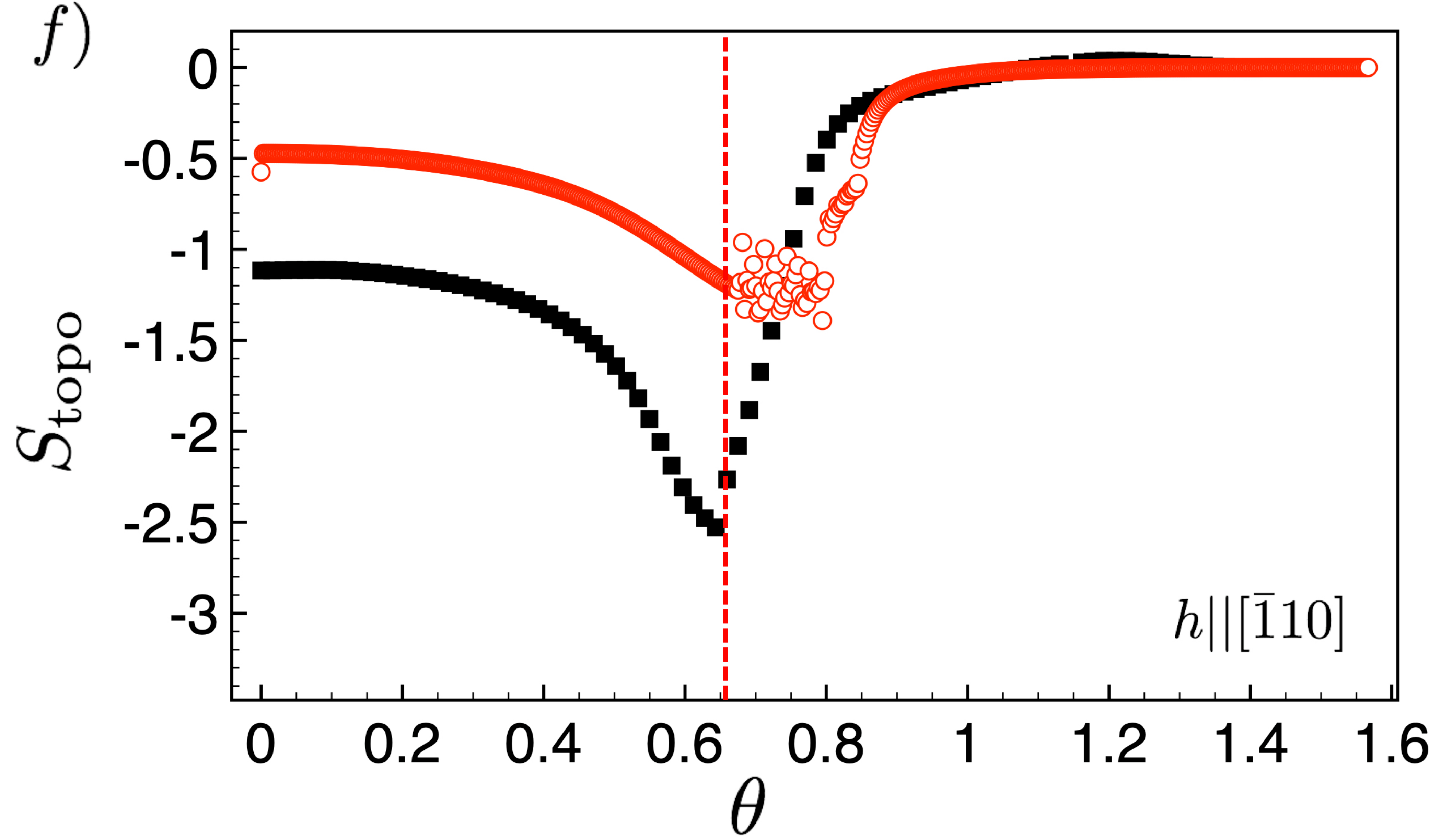}
\caption[]{Energies of the lowest lying excitations of the AF case of Eq. (\ref{kitaev}) relative to the ground state energy vs the field strength parameter $\theta$ for a field along (a) $[111]$, (b) $[001]$, and (c) $[\bar{1}10]$ for a 24-site cluster. (d)$-$(f) show the topological entanglement entropy $S_{\text{topo}}$ for fields along $[111]$, $[001]$, and $[\bar{1}10]$, respectively, for 18-site (red) and 24-site (black) clusters. The vertical red lines mark discontinuities in $S_{\text{topo}}$ that corroborate changes in the eigenvalue spectrum. (d)$-$(e) show two distinct transitions from a gapped non-Abelian Kiatev QSL at low field, to a gapless QSL phase for intermediate fields, and finally to a gapped polarized phase at high fields. (f) is consistent with a single transition, given finite-size limitations.}
\label{fig1}
\end{center}
\end{figure*}

Motivated by recent experiments on $\alpha$-RuCl$_{3}$ in an externally applied magnetic field \cite{Leahy2017, Sears2017, Wolter2017, Baek2017, Do2017, Zheng2017, Banerjee2018, Hentrich2018, Seifert2018}, and their analyses using extended magnetic models \cite{Yadav2016, Janssen2016, Janssen2017, Chern2017, Gohlke2017, Winter2018}, we investigate the quantum phase transitions of the pure Kitaev model as a function of an externally applied magnetic field. Isolating the Kitaev term, which is responsible for imparting to candidate materials their exotic nature, gives us greater insight into what might underlie these materials' salient features under an applied field. Using exact diagonalization (ED) on up to 24-site clusters, we observe unambiguous signatures in the energy spectrum, topological entanglement entropy (TEE), and the dynamical local spin-spin correlations. Our main results are as follows: (i) We find phase transitions from a gapped Kitaev QSL to an intermediate gapless QSL, to a partially polarized phase with increasing field along either $[111]$ or $[001]$, for antiferromagnetic (AF) Kitaev interactions. These are deduced from the increased density of states (DOS) in the energy spectrum, as well as anomalies present in the TEE as a function of field strength. (ii) Importantly, the intermediate gapless QSL phase is considerably reduced for fields along [$\bar{1}$10]. A direct transition between the non-Abelian Kitaev QSL and polarized phases cannot be ruled out for fields pointing near the [$\bar{1}$10] direction. (iii) The local dynamical response along $[111]$ shows a plethora of modes for intermediate values of the field strength, that are considerably reduced upon increasing the strength of the field towards the polarized limit, resulting in a clearly discernible beating pattern between a few modes of comparable strength and energy. (iv) The behavior of the average of plaquette operators $\langle W_{p}\rangle$ and its dependence on field strength and orientation provides a useful diagnostic of the non-Abelian Kitaev QSL, intermediate, and polarized phases, and reveals differences between the ferromagnetic (FM) and AF cases.

{\it Model and approach.}
We use exact diagonalization on up to 24-site clusters with periodic boundary conditions (see Supplemental Material \cite{supp2018}). We define the Hamiltonian:
\begin{equation}\label{kitaev}
H=\sum_{\alpha}\left(J_{\alpha}\sum_{\langle j,k\rangle_{\alpha}}\sigma_{j}^{\alpha}\sigma_{k}^{\alpha}\right)-\vec{h}\cdot\sum_{j}\vec{\sigma}_{j},
\end{equation}
and resort to the following parametrization
\begin{eqnarray}\label{param}
J_{x}&=&J_{y}=J_{z}=J=\pm\cos\theta,\\
\vec{h}&=&\frac{1}{\sqrt{2\lambda^{2}+1}}(\lambda,\lambda,1)\sin\theta,
\end{eqnarray}
with $\theta=\tan^{-1}(|\vec{h}|/J)$, $0\leqslant\theta\leqslant\pi/2$, $0\leqslant\lambda\leqslant1$, and the positive (negative) case corresponding to AF (FM) near-neighbor interactions along respective bonds $\alpha\in\{x,y,z\}$. We only consider the isotropic point interaction space and vary its strength $\theta$, and along varying orientations ranging from $[111]$ for $\lambda=1$ to $[001]$ for $\lambda=0$. We also consider the special case of a field along $[\bar{1}10]$, lying within the plane of the honeycomb lattice. For each of these field directions, we obtain the energy spectrum, and the TEE as a function of $\theta$ using the Kitaev-Preskill construction \cite{Kitaev2006b}.

In order to connect with experiments, we calculate the dynamical correlations $S_{jk}^{\alpha\alpha}(t,\theta,\lambda)=\langle0_{\theta,\lambda}|\sigma_{j}^{\alpha}(t)\sigma_{k}^{\alpha}(0)|0_{\theta,\lambda}\rangle,$ 
where $|0_{\theta,\lambda}\rangle$ is the field strength and field orientation-dependent ground state of Eq. \ref{kitaev}, and $\sigma_{j}^{\alpha}(t)=e^{iHt}\sigma_{j}^{\alpha}(0)e^{-iHt}$. 
We also calculate the field-dependent on-site time Fourier transform,
\begin{equation}\label{tfour}
\begin{aligned}
&S_{jk}^{zz}(\omega, \theta,\lambda)=\int_{-\infty}^{\infty}S_{jk}^{zz}(t,\theta,\lambda)e^{i\omega t}\,dt\\
&=\sum_{n}\langle0_{\theta,\lambda}|\sigma_{j}^{z}(0)|n\rangle\langle n|\sigma_{k}^{z}(0)|0_{\theta,\lambda}\rangle\,
\delta(\omega+E_{0}-E_{n}),
\end{aligned}
\end{equation}
where $n$ is the energy quantum number indexing the various eigenenergies and eigenstates of $H$. Below, we restrict our calculations to the case of $j=k$ and $\alpha=z$.

{\it Signatures of field-driven quantum phase transitions.}
For AF couplings of the Kitaev honeycomb model under an externally applied field, recent findings reveal a rich phase diagram, beyond the perturbative result \cite{Zhu2017,Gohlke2018,Lu2018}. Our aim is to connect features in the energy spectrum, as a function of $\theta$, that indicate phase transitions, with specific signatures in the dynamical correlation functions, allowing us to make testable predictions  
for inelastic neutron scattering and optical-pump terahertz-probe spectroscopy.

{\it Energy spectrum and TEE.} 
For the AF case, for a 24-site cluster and a field along $[111]$ [Fig. \ref{fig1}(a)], the salient feature in the excitation spectrum is the dramatic increase in the DOS at lower $\Delta E$ within the broad range $0.63\leq\theta\leq0.88$. A qualitatively similar feature occurs for a field along $[001]$ [Fig. \ref{fig1}(b)], where the lowest excitations also form a broad continuum within the range $0.44\leqslant\theta\leqslant0.82$. 
For either of these field orientations we label the lower (higher) bound of these high DOS regions $\theta^{*}$ $(\theta^{**})$.

We see corresponding discontinuities in the TEE curves at respective $\theta^{*}$ and $\theta^{**}$ values for each of the two field orientations [black squares in Fig \ref{fig1}(d) and \ref{fig1}(e)]. Taken together, the critical values identify the locations of phase transitions to and from an intermediate quantum phase for these two field orientations. 

Now, the $\theta=0$ limit with considerably lower DOS at low $\Delta E$ is known to be a gapless QSL in the thermodynamic limit. We therefore conclude that the enhanced DOS in the region between $\theta^{*}$ and $\theta^{**}$, as well as the discontinuities observed in the TEE at these latter points, indicate the existence of a second, different, gapless QSL sandwiched between the gapped non-Abelian Kitaev QSL and the polarized phase expected for higher values of $\theta$, for each of the two field orientations $[111]$ and $[001]$.

The gapped QSL phase is clearly discernible as having a negative value of TEE, while the polarized phase has TEE close to zero, as expected. For gapless systems, due to a $\log L$ correction in the area law, the Kitaev-Preskill construction can no longer be used to ascertain QSL behavior. We continue to use the Kitaev-Preskill construction for intermediate values of $\theta$ however, not to get an accurate value of the TEE, but to identify phase transitions to and from the intermediate gapless phase. 

The comparison of TEE on two cluster sizes, 18 and 24 sites [Fig. \ref{fig1}(d)-(f)], shows the shift in the locations of the jump discontinuities of TEE around $\theta^{*}$, towards lower $\theta$, with increasing cluster size.
Finally, for a field along $[\bar{1}10]$ [Fig. \ref{fig1}(c) and 1(f)], while we observe an increase in the DOS (and a corresponding discontinuity in TEE) near $\theta\approx0.66$, there does not appear to be a range in $\theta$ supporting a higher DOS as is true of the other two field orientations along $[111]$ or $[001]$. There is also a lack of a clear discontinuity in the TEE which would suggest an upper bound for any presumptive intermediate phase. While we do not rule out the possibility of the presence of an intermediate phase for this particular orientation of the external field, within the numerics reported here we cannot guarantee its presence either.

\begin{figure}
\begin{center}
\includegraphics[scale = .44]{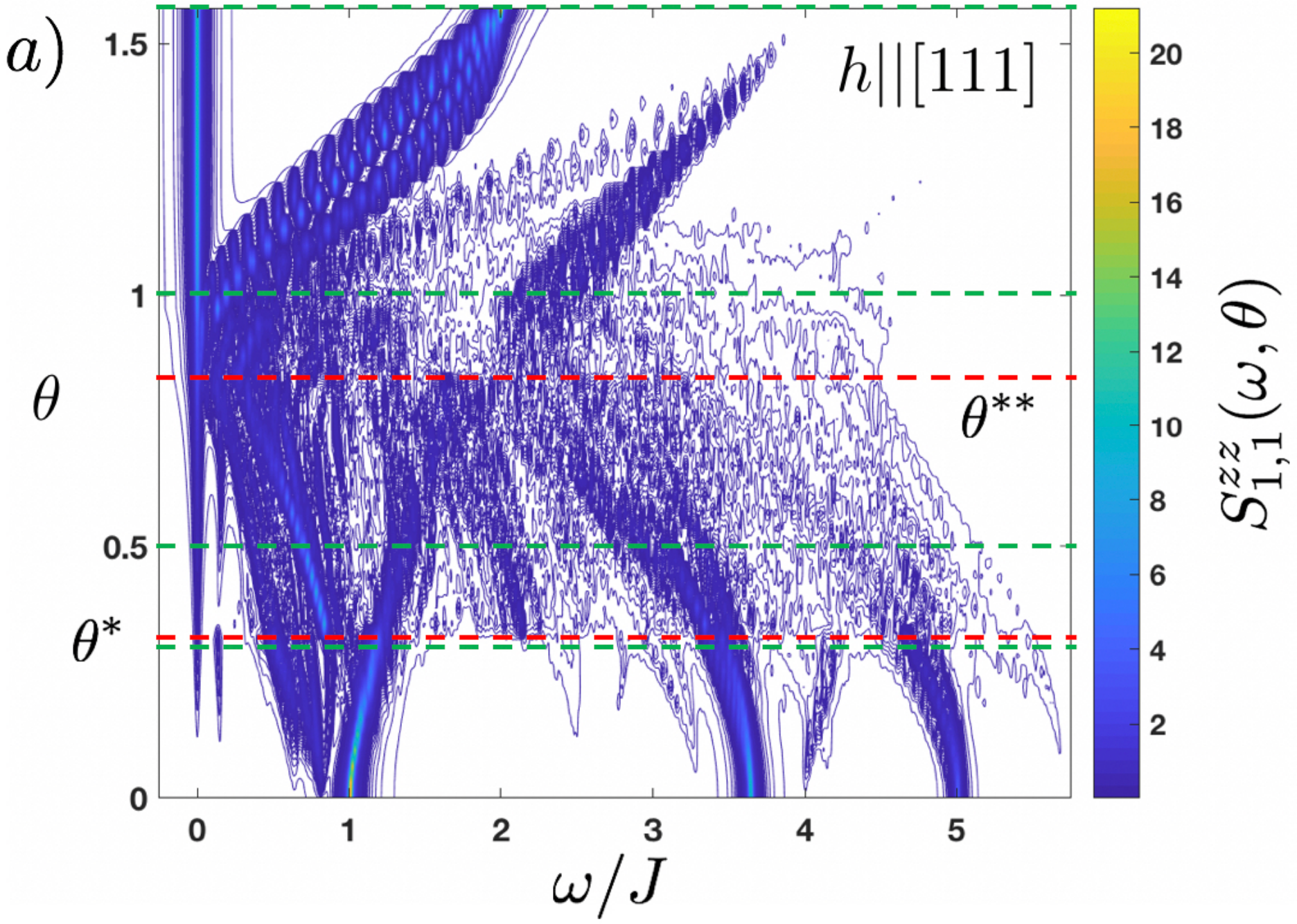}
\includegraphics[scale = .20]{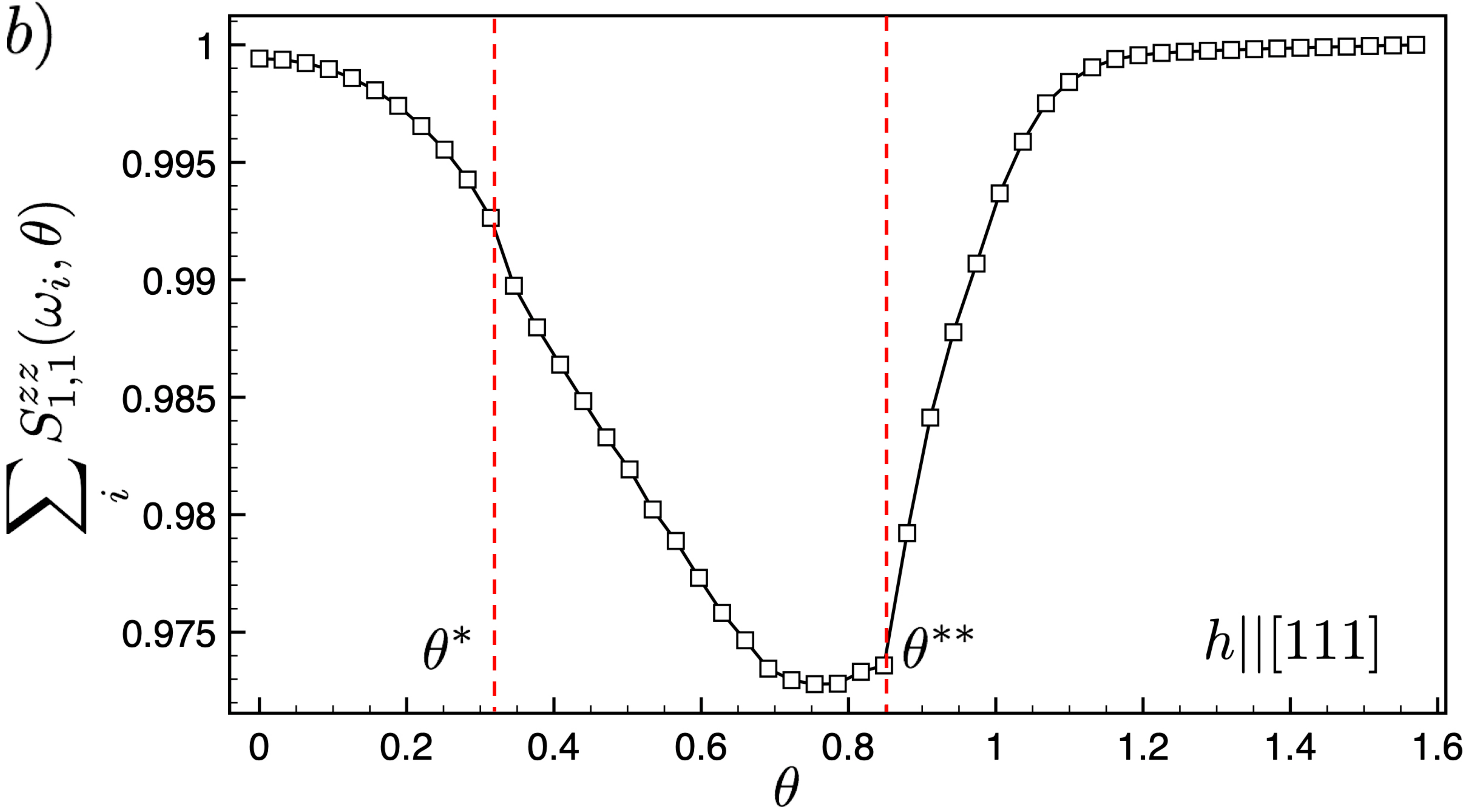}
\caption[]{The on-site dynamical response for the AF case and the normalized integrated intensity at constant $\theta$ for $h\parallel [111]$. Cuts along the green lines in (a) are shown in the Supplemental Material \cite{supp2018}. Red lines  correspond to the locations of phase transitions, shown here for a 16-site cluster.}  
\label{fig2}
\end{center}
\end{figure}

\begin{figure*}
\begin{center}
\begin{tabular}{cccc}
\includegraphics[scale = .143]{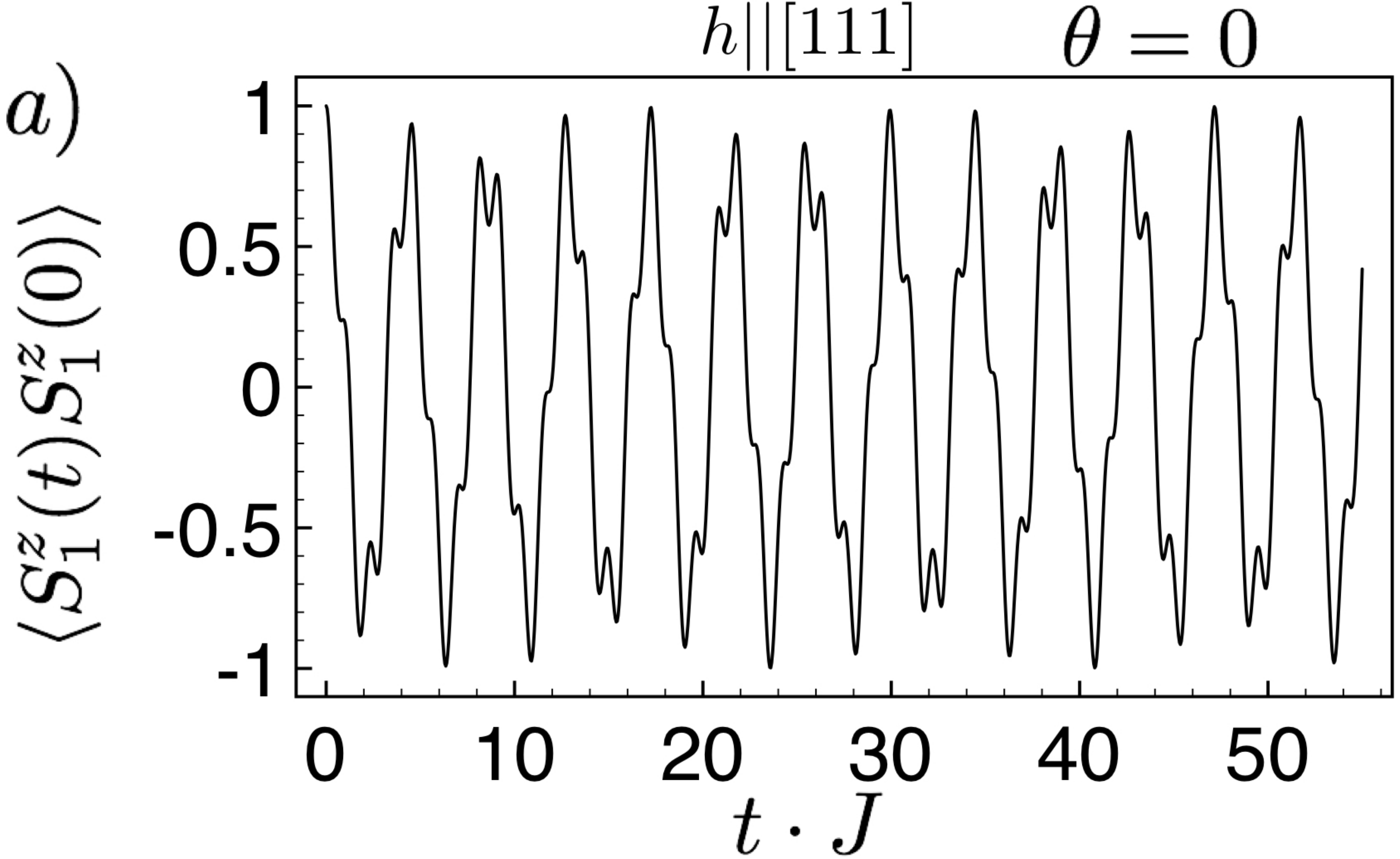}&
\includegraphics[scale = .155]{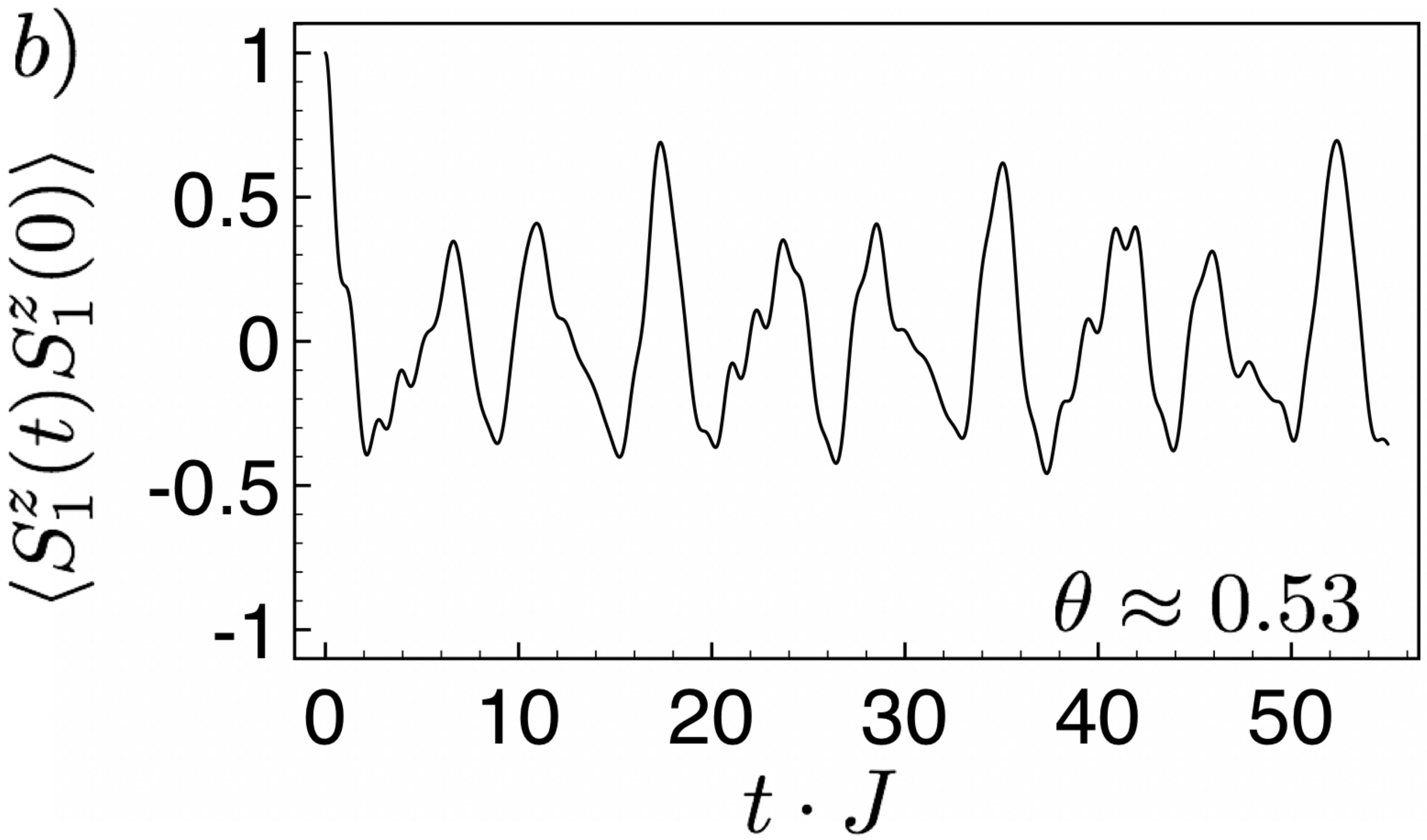}&
\includegraphics[scale = .155]{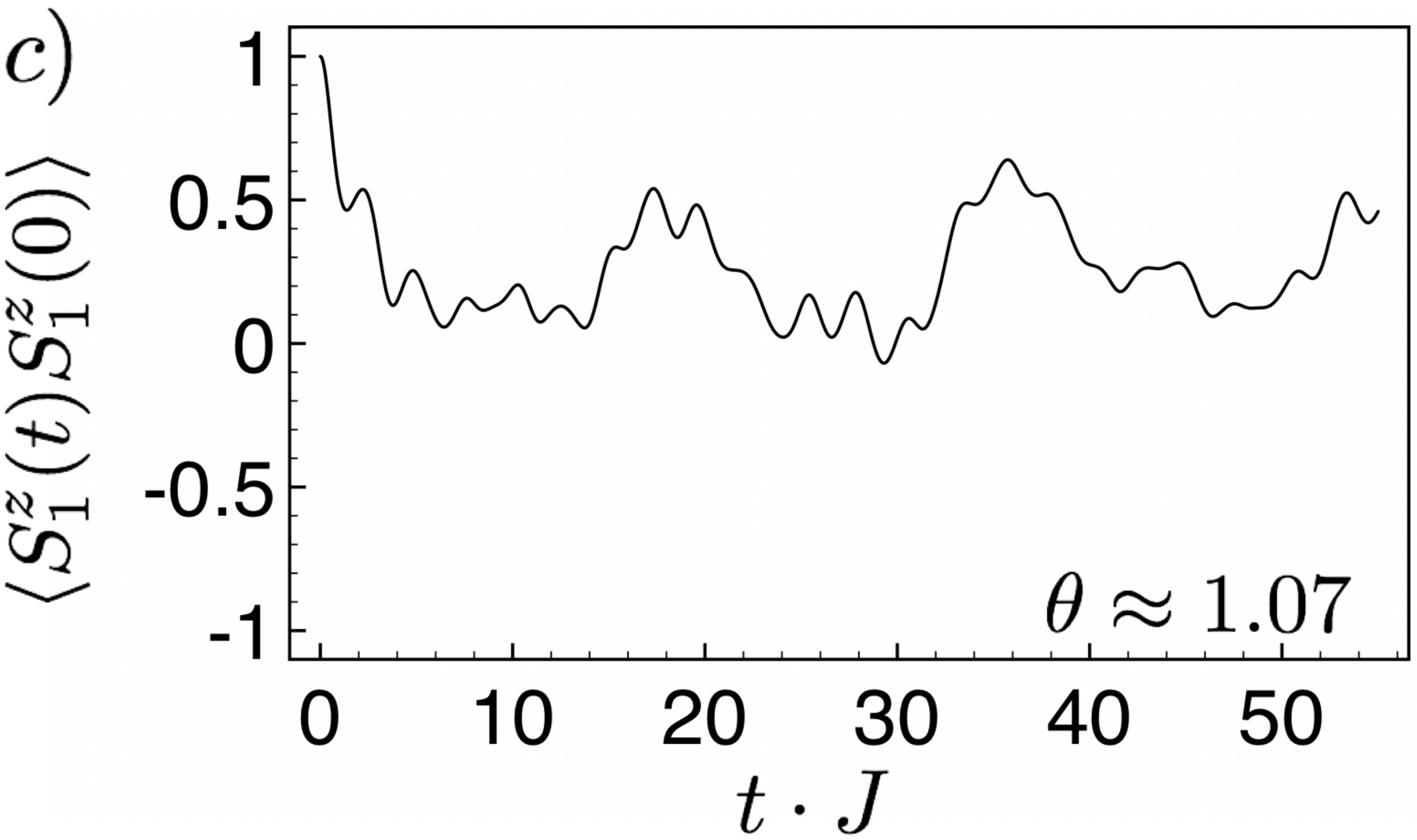}&
\includegraphics[scale = .155]{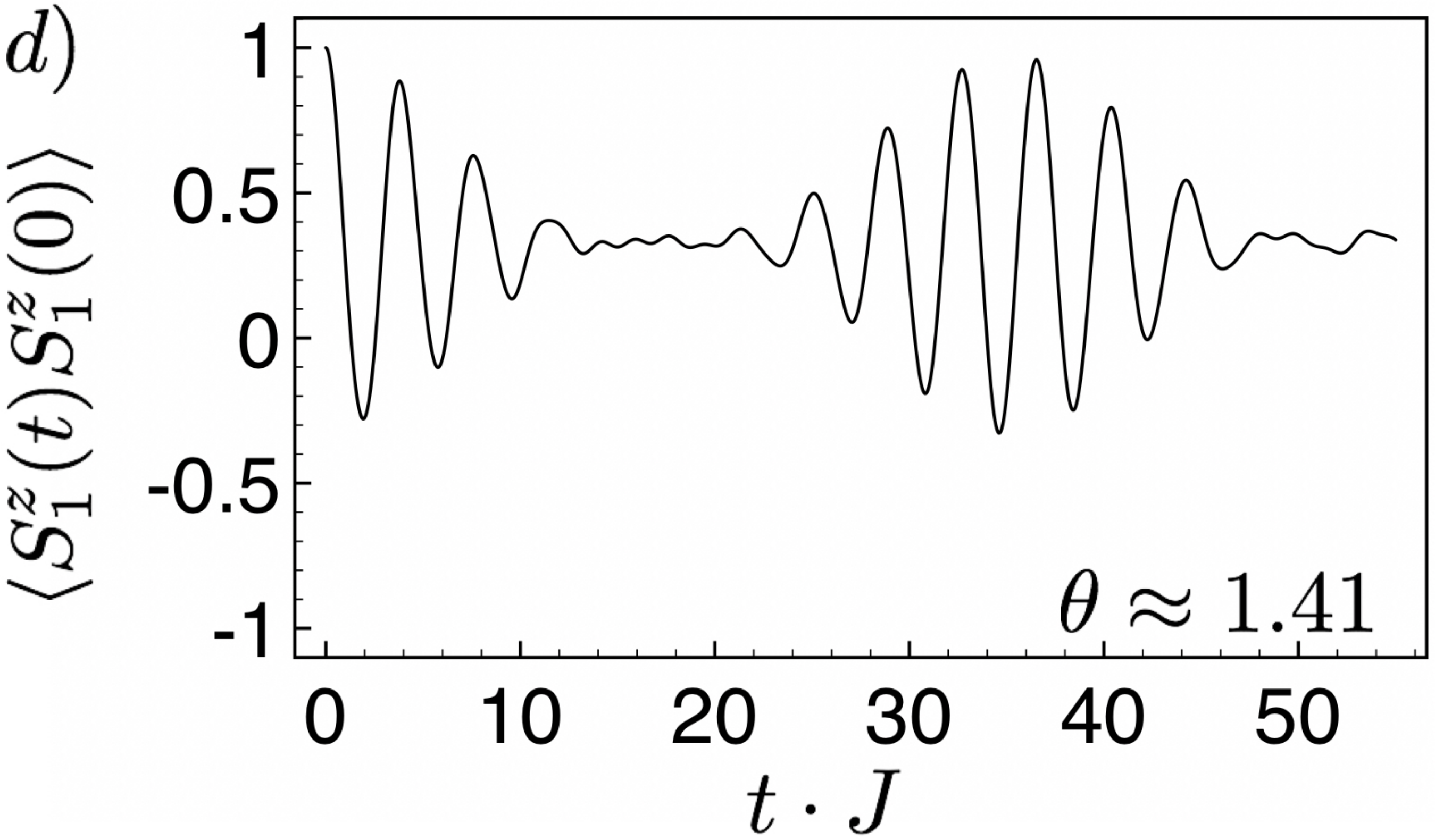}\\
\end{tabular}
\includegraphics[scale = .6]{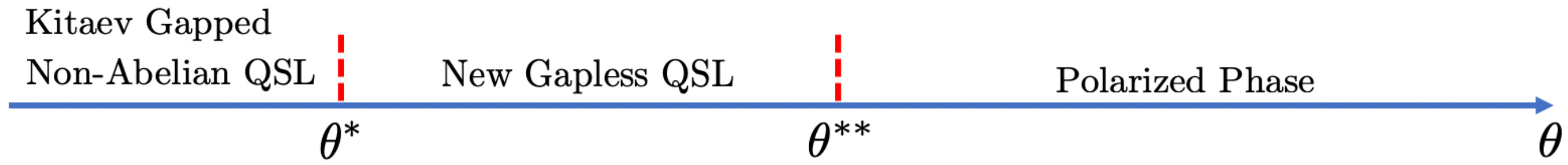}
\caption[]{The on-site dynamical spin-spin correlations, for AF exchange, for a field along $[111]$, with $\theta^{*}\approx0.32$ and $\theta^{**}\approx0.85$.}
\label{fig3}
\end{center}
\end{figure*}

{\it Dynamical correlations for field along [111].} 
The evolution of the on-site time Fourier transform Eq. (\ref{tfour}) with field strength [Fig. \ref{fig2}(a)] shows sharp and intense modes that follow independent trajectories
for $\theta< \theta^{*}\approx0.32$.
At $\theta^{*}$, discontinuities in the trajectories of various modes are observed, and beyond that within $\theta^{*}<\theta<\theta^{**}\approx0.85$, there is a drastic decrease in the intensity of modes which now form a featureless continuum across the entire spectrum [see Fig. \ref{fig2}(b)]. Just above $\theta^{**}$, a featureless continuum of comparably intense modes having no sharp or well-defined peaks forms into well-defined modes of considerable intensity at lower energies as $\theta$ is increased. The trajectories of these latter modes tend towards higher $\omega$ with increasing $\theta$, persisting up to $\theta=\pi/2$ where they converge and become the most dominant mode at $\omega/J\approx2$. The nature of the excitation corresponding to this latter mode consists of a single spin flip about the completely polarized state along $\hat{z}$.

\begin{figure*}[htb!]
\begin{center}
\includegraphics[scale = .2]{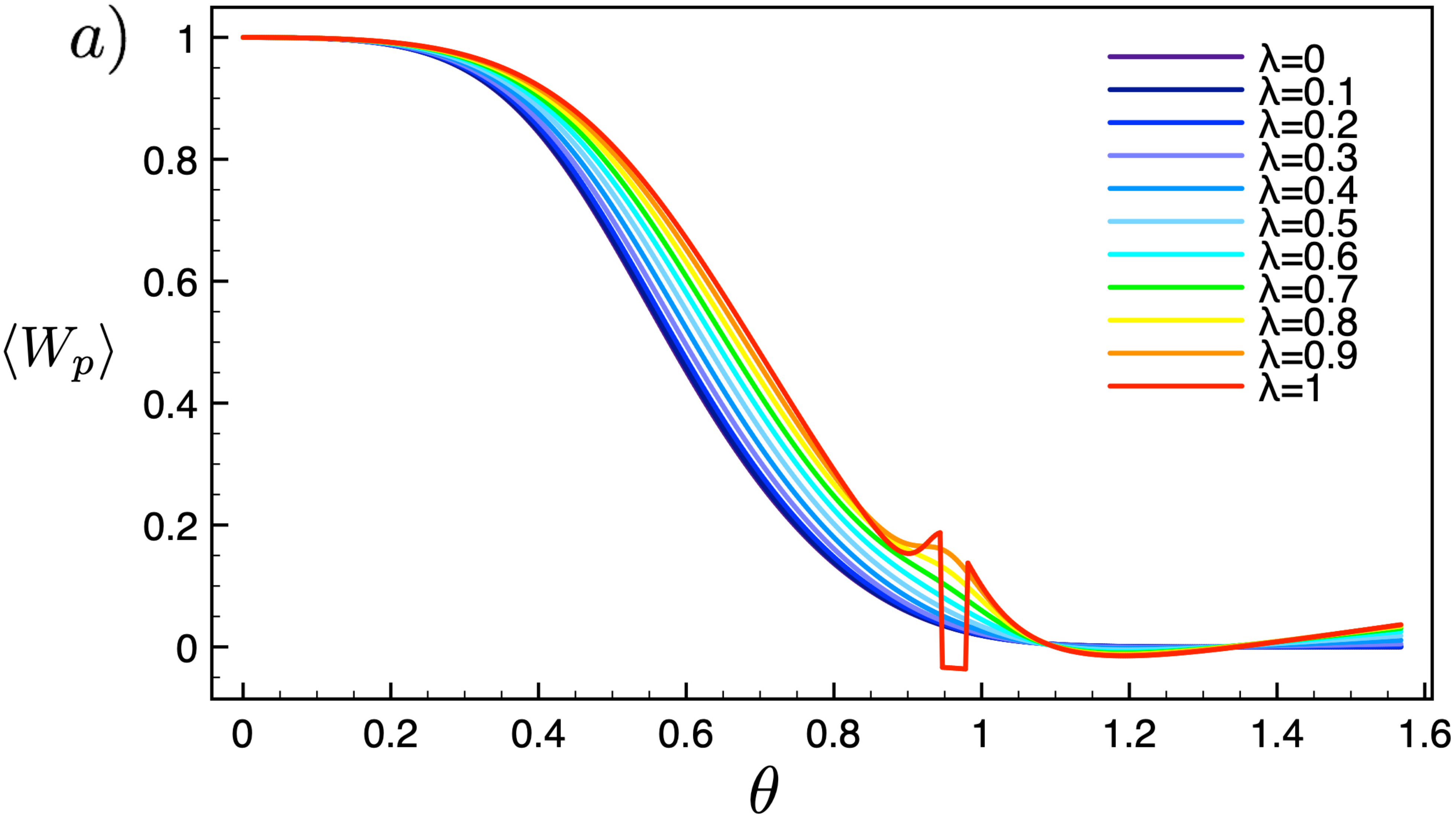}\qquad\quad
\includegraphics[scale = .2]{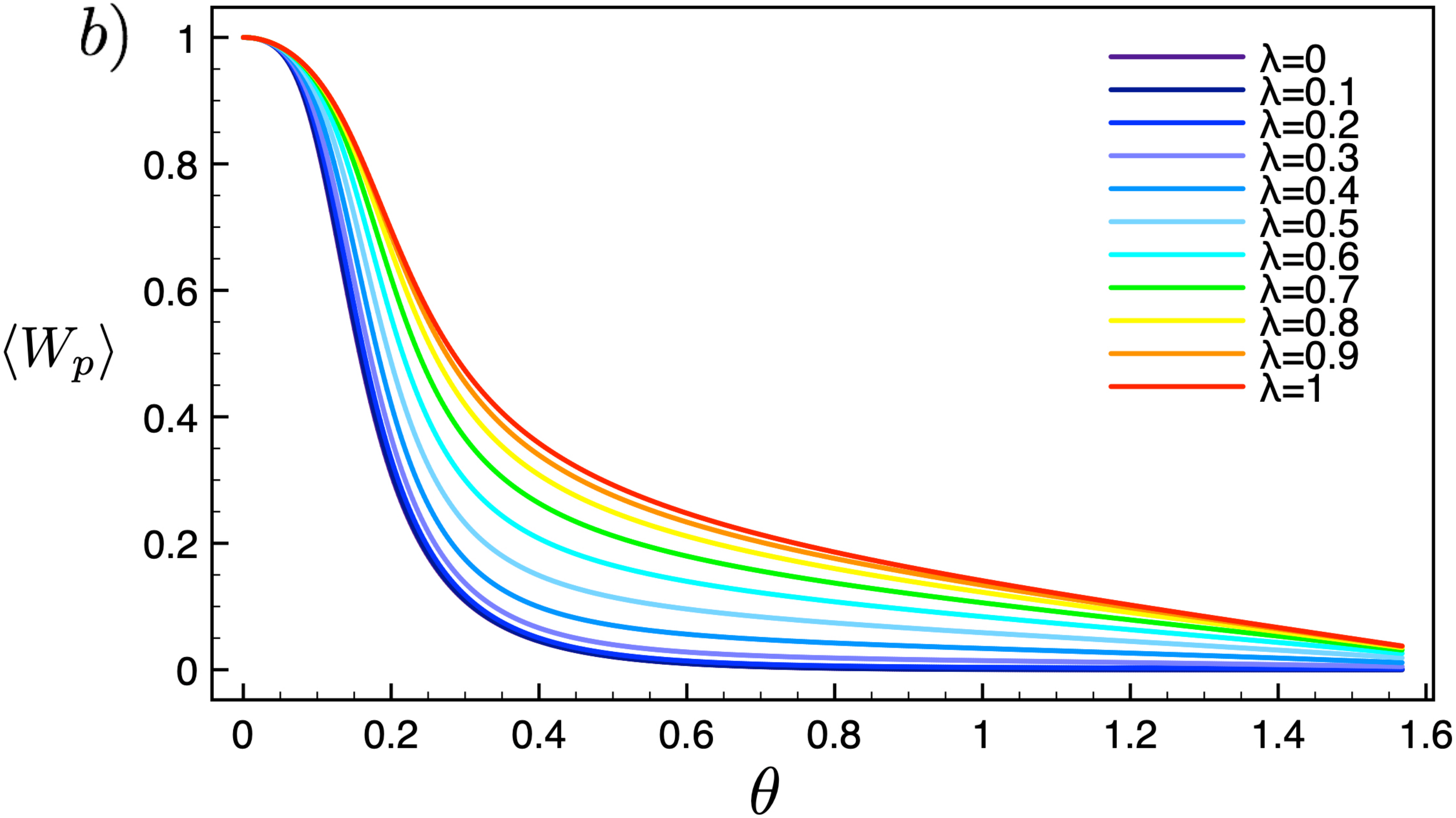}
\caption[]{Average of the plaquette operator $\langle W_{p}\rangle$, for (a) AF and (b) FM cases, as a function of both the field strength parameter $\theta$ and field orientation parameter $\lambda$. For AF interactions, (a) shows a clear intermediate region. For FM interactions, a rather abrupt transition between the Kitaev QSL and the polarized phase is observed in (b). In each case, the non-Abelian Kitaev QSL phase increases as a function of $\lambda$, and is more pronounced in the AF case.}
\label{fig4}
\end{center}
\end{figure*}

{\it Real-time dynamics.}  
The local dynamical spin-spin correlations for $h\parallel [111]$ evolve from the behavior at $\theta=0$ with the waveform in Fig. \ref{fig3}(a) arising from the superposition of a dominant low-frequency mode with a pair of lower-intensity, high-frequency modes [see the cut along the horizontal axis in Fig. \ref{fig2}(a) above, and in the Supplemental Material \cite{supp2018}]. Within the gapped Kitaev QSL ($0\leqslant\theta\lesssim\theta^{*}$), the wave forms are characterized by a large amplitude and long wavelength, modulated by small oscillations. The low-energy mode responsible for the long wavelength in $S_{11}^{zz}(t,\theta)$ reflects the two-flux energy gap, while the high-energy modes responsible for the small oscillations reflect single and multiparticle processes \cite{Knolle2014, Knolle2015}. 

Wave forms of $S_{11}^{zz}(t,\theta)$ lying within $\theta^{*}<\theta<\theta^{**}$ [Fig. \ref{fig3}(b)] are relatively featureless because of a broad continuum of comparably intense modes across a wide range of energies [see the horizontal cut along $\theta\approx0.50$ in Fig. \ref{fig2}(a) above, and in the Supplemental Material \cite{supp2018}]. This featureless character of $S_{11}^{zz}(t,\theta)$ persists above $\theta^{**}$ [Fig. \ref{fig3}(c)]. Within the partially polarized phase, the frequency response is once again characterized by a few sharp and highly intense modes, which depict a distinct beat feature, or wave packet, due to interference between a small number of sharp, comparably intense, and energetic modes [Fig. \ref{fig3}(d)]. We believe this beat feature, which may be measurable using pump-probe THz spectroscopy on candidate materials, is a telltale signature of the onset of a QSL phase from a polarized phase.

{\it Field strength and orientation dependence of plaquette flux.}
The plaquette operator averaged over the entire lattice, with respect to the ground state, is
\begin{equation}\label{plaqavg}
\langle W_{p}\rangle=\frac{1}{4}\sum_{i=1}^{4}\langle W_{p_{i}}\rangle=\frac{1}{4}\sum_{i=1}^{4}\left\langle \prod_{j\in p_{i}}\sigma_{j}^{\alpha(j)}\right\rangle,
\end{equation}
in this case over four plaquettes $p_{i}$, with bonds $\alpha(j)$ emanating from site $j$ away from the interior of the plaquette. 

The eigenvalues of $W_{p_{i}}$ are $\pm1$, and $[W_{p_{i}},H]=0$ when $\theta=0$. In this limit, the ground state lies within the $W_{p_{i}}=1$ block for all $p_{i}$ \cite{Lieb1994}, and an excitation corresponding to $W_{p_{i}}=-1$ indicates the presence of a flux. With deviation $\theta\ne0$, $\langle W_{p}\rangle\approx 1$ despite $[W_{p_{i}},H]\ne0$, corresponding to the non-Abelian Kitaev QSL phase.

We see that, in agreement with earlier studies \cite{Zhu2017, Gohlke2018}, this phase appears to extend further in applied field $\theta$ in the AF case than in the FM case (Fig. \ref{fig4}). 

What is different here is our finding that the non-Abelian Kitaev QSL phase in the AF case [Fig. \ref{fig4}(a)] expands as the field is rotated from the $[001]$ to the $[111]$ direction.

In the FM case [Fig. \ref{fig4}(b)] there is a sharper decrease in $\langle W_{p}\rangle$, vs $\theta$, than in the AF case, suggesting a direct transition between the non-Abelian Kitaev phase and the polarized phase for $\lambda\approx0$, and possibly also as $\lambda$ increases toward unity.

{\it Conclusions.}
We have expanded the exploration of QSLs from thermodynamics and spectroscopy to probing the nature of fractionalized excitations directly in the real-time dynamics using pump-probe THz spectroscopy.
In these experiments, the pump excites photocarriers in the system, and the THz probe pulse measures the photoconductivity as a function of time \cite{Jiang2011,Zhang2014}. Using an exchange coupling of about 5 meV, for either the AF or FM case, we expect signatures of fractionalization to appear at time scales lying within the range $10^{-13}\text{ s} <t<10^{-12}\text{ s}$, or for frequencies in the 1$-$10 THz regime. We bring to light facts pertaining to the field strength and orientation dependence of phases exhibited by the FM and AF Kitaev honeycomb models under an externally applied magnetic field. At this stage, the nature of the intermediate gapless QSL phase for AF Kitaev interactions requires further investigation. 

{\it Acknowledgements.} We thank Kyusung Hwang, Nirav Patel, Kyungmin Lee, and Yuan-Ming Lu for their helpful comments and discussions. We acknowledge the support of the DOE-BES Grant No. DE-FG02-07ER46423.

\end{document}